\documentclass{MolSpaLab}
\usepackage{graphicx}

\begin{document}
\TitreGlobal{Cosmic Dust Near and Far}
%%-----------------------------
%%      the top matter
%%-----------------------------
\title{Interstellar extinction and abundances of 
simple diatomic molecules}
\author{\FirstName T. \LastName Weselak}
\address{Department of Physics, Kazimierz Wielki University,  Weyssenhoffa 11,
85-072 Bydgoszcz, Poland}
% un-comment if additional authors
\author{\FirstName J. \LastName Kre{\l}owski}
\address{Center for Astronomy, Nicolaus Copernicus
University,
Gagarina 11, Pl-87-100 Toru{\'n}, Poland
}

\runningtitle{Interstellar extinction and abundances of 
simple diatomic molecules}
\setcounter{page}{1}

\maketitle
\begin{abstract}
We analyze mutual relations between interstellar extinction E(B-V) 
and column densities of diatomic molecules such as 
CN, CH and CH cation observed along translucent sightlines. 
Observational material acquired using five 
echelle spectrographs, situated in both northern and southern 
hemispheres, was used in this study. While column densities of 
CH, CH$^{+}$ correlate reasonably tightly with E(B-V) the similar
relation built for CN shows a large scatter. 
This result suggests that CH and CH$^{+}$ molecules are 
quite well spatially correlated to interstellar dust while CN requires
another conditions to be formed. 
The CH molecule is closely related to H$_{2}$ as it was 
demonstrated already; the fact is interesting since H$_{2}$ is believed
to be formed on grain surfaces rather than in the gas phase.
\end{abstract}

\section{Introduction}

The aim of this project is:

\begin{itemize}
    \item To determine column densities of simple diatomic\\ 
molecules toward new 80 targets in our database

  \item To check relations between abundances of CH, CH$^{+}$, 
CN molecules and E(B-V) basing on statistically meaningful sample 
of objects

  \item To find out the new relation between column densities 
of CH and H$_{2}$ molecules based on new results concerning CH molecule
(abundances of molecular hydrogen are published by 
Savage et al. 1977, Rachford et al. 2002).
\end{itemize}

\section{The Observational Material}
\label{section1}
Our observing material consists of spectra 
of 168 stars containing CN and/or CH and CH$^{+}$ 
features. It was acquired using five echelle spectrometers:
\begin{itemize}

  \item MAESTRO fed by the 2-m telescope of the Observatory 
at Peak Terskol (see http://www.terskol.com\ /telescopes/3-camera.htm)
  \item Feros spectrograph, fed with the 2.2m ESO 
telescope in Chile\\ (see http://www.ls.eso.org/lasilla/sciops/2p2/E2p2M/FEROS/)
  \item fiber-fed echelle spectrograph installed at 
1.8-m telescope of the Bohyunsan Optical Astronomy Observatory (BOAO) in South Korea
  \item HARPS spectrometer, fed with the 3.6m ESO 
telescope in Chile\\ (see 
http://www.ls.eso.org/lasilla/sciops/3p6/harps/)
  \item UVES spectrograph at Paranal in Chile. 
For more information see:\\ 
http://www.eso.org/sci/facilities/paranal/instruments/uves

\end{itemize}
\noindent
To obtain column densities of diatomic molecules we used:
\begin{itemize}
\item in the case of CN features with EW$>$14m\AA\ (CN features are saturated)
we used the method of Roth \& Meyer (1995).  
  \item when CH line of A-X system  was saturated 
we used the sum of column densities obtained from 
unsaturated 3886 and 3890 lines of B-X system.
  \item in the case of saturated CH$^{+}$ A-X (0, 0) band 
at 4232 \AA\ we used unsaturated (1, 0) band centered near 3957 \AA.
\end{itemize}

\noindent
Oscillator strengths in the case of each transition are given 
in Table 1.\\

\noindent
{\small{
{\bf{Table 1}}: Adopted molecular parameters. References: 
1~--~Gredel et al. (1993), 2~--~Weselak et al. (2009a), 3~--~van Dishoeck and Black (1986),
4~--~Roth and Meyer (1995).

\begin{tabular}{lrclclcccc}
\hline\hline
& & & & & & & & &\\
Species &   Vibronic band               &   Rotational lines        &   Position   &Ref.   &   f-value &Ref.    \\
& & &[\AA] &  & & & & &\\
\hline
& & & & & & & & &\\
CH  &   A$^{2}\Delta$ -- X$^{2}\Pi$ (0, 0)		&   R$_{2e}$(1) + R$_{2f}$(1)     &   4300.3132   &1  &   0.00506  &3 \\
    &   B$^{2}\Sigma^{-}$ -- X$^{2}\Pi$ (0, 0)  	&   Q$_{2}$(1)+$^{Q}$R$_{12}$(1)  &   3886.409    &1  &   0.00320 &1  \\
    &                  			 (0, 0)  	&   $^{P}$Q$_{12}$(1)          	  &   3890.217    &1  &   0.00210 &1  \\
CH$^{+}$&   A$^{1}\Pi$ -- X$^{1}\Sigma^{+}$ (0, 0)   	& R(0)               		  &   4232.548    &2  &   0.00545 &2  \\
    &   				(1, 0)          &   R(0)                	  &   3957.689    &2  &   0.00342 &2  \\
CN  &  B$^{2}\Sigma^{+}$ -- X$^{2}\Sigma^{+}$ (0, 0)    &   R(0)                          &   3874.608    &4  &   0.0342  &4  \\
    &                                                   &   R(1)                          &   3873.998    &4  &   0.0228  &4 \\
    &         						&   P(1)     			  &   3875.763    & 4 &   0.0114  &4\\
    
\hline
\end{tabular}
}}

\section{Results}
\label{section2}

\begin{itemize}
   \item The interstellar reddening correlates better 
 with CH than CH$^{+}$ molecule. 
 A very poor relation is observed in the case of CN.
   \item Certain objects (HD 27778, 147889, 154368, 161056, 
 204827, 208501) demonstrate high abundances of 
 diatomic molecules in relation to E(B-V) \\  -- 
 CH\&CN abundant clouds
   \item HD 34078 is untypical object in case of which 
 CH and CH$^{+}$ molecules are overabundant in relation to E(B-V) 
 -- CH\&CH$^{+}$ abundant cloud
 \item The relation between column densities of CH and H$_{2}$
 molecules is very tight (see also results of Danks, Federman and Lambert 1984, 
 Weselak et al. 2004). 
 New relation with correlation coefficient equal to 0.95 is 
obtained on the basis of 53 stars.
\end{itemize}

\begin{figure*}[ht!]

\includegraphics[width=6.5 cm]{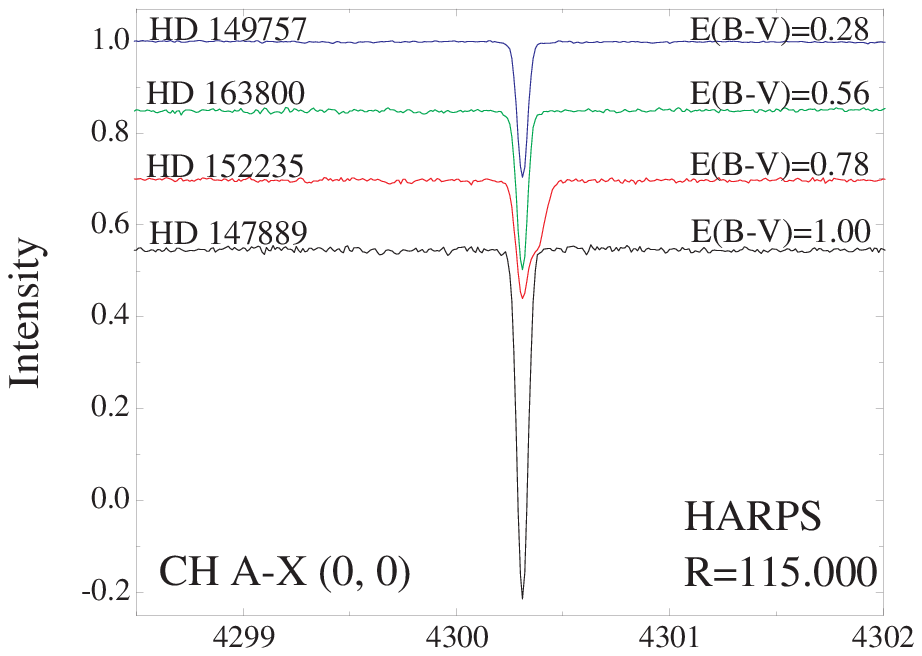}
\includegraphics[width=6.5 cm]{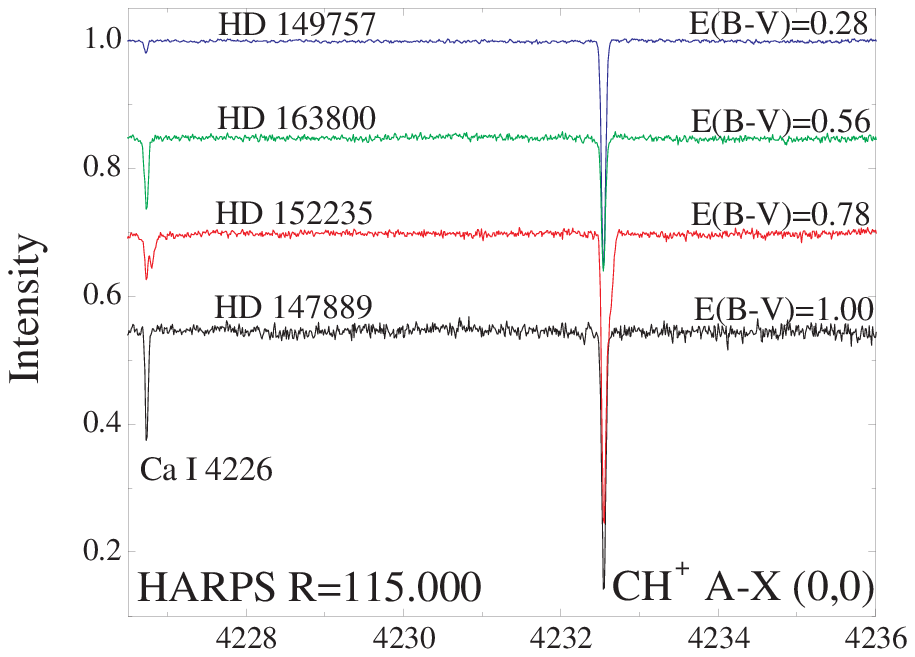}
\caption{ The CH A-X system at 4300 \AA\ in spectra 
of four targets with increasing reddening
(at the left). At the right panel we present 
CH$^{+}$ A-X (0, 0) band at 4232 \AA\ seen in the
spectra of the same stars.
}
\label{fig}
\end{figure*}

\begin{figure*}[ht!]
\includegraphics[width=6.5 cm]{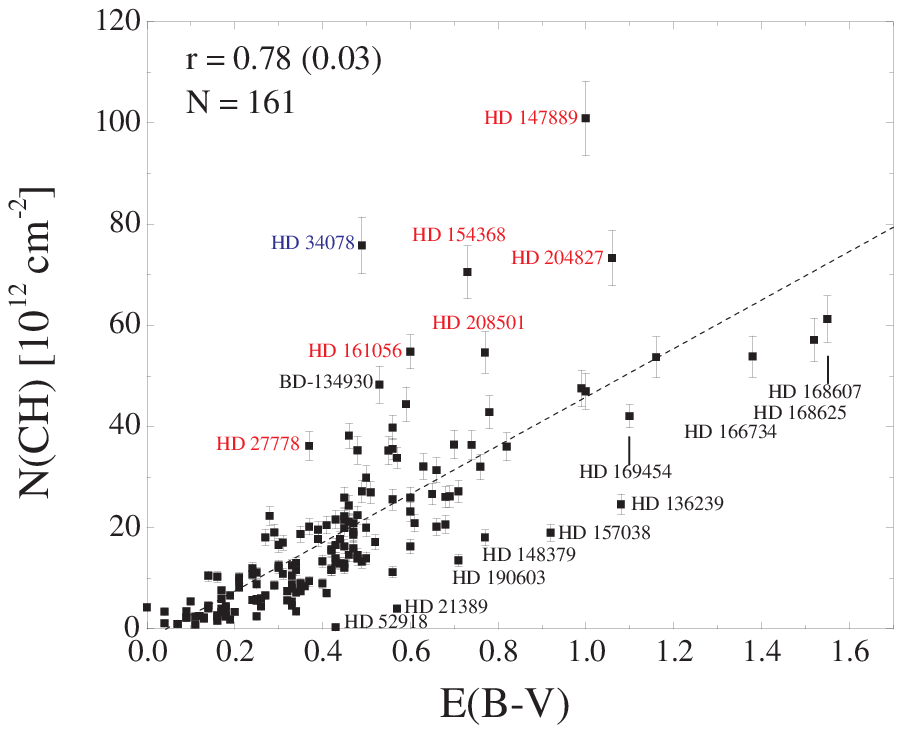}
\includegraphics[width=6.5 cm]{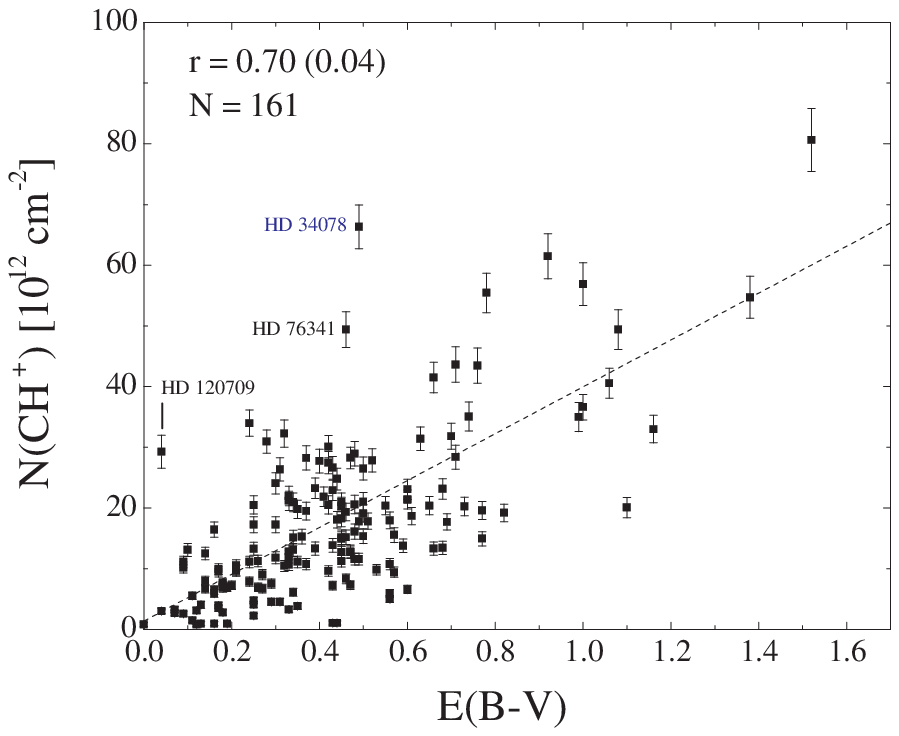}
\caption{ 
Correlation plots between column densities 
of CH, CH$^{+}$
versus E(B-V). The best relation is seen in the case of CH molecule 
which is apparently also a very good H$_{2}$ tracer. 
Objects with overabundant CH and CN in relation to E(B-V) are 
marked as red and CH\&CH$^{+}$ object HD 34078 in spectrum of 
which CH and CH$^{+}$ are highly abundant -- with blue.}
\label{fig}
\end{figure*}

\begin{figure*}[ht!]
\includegraphics[width=6.5 cm]{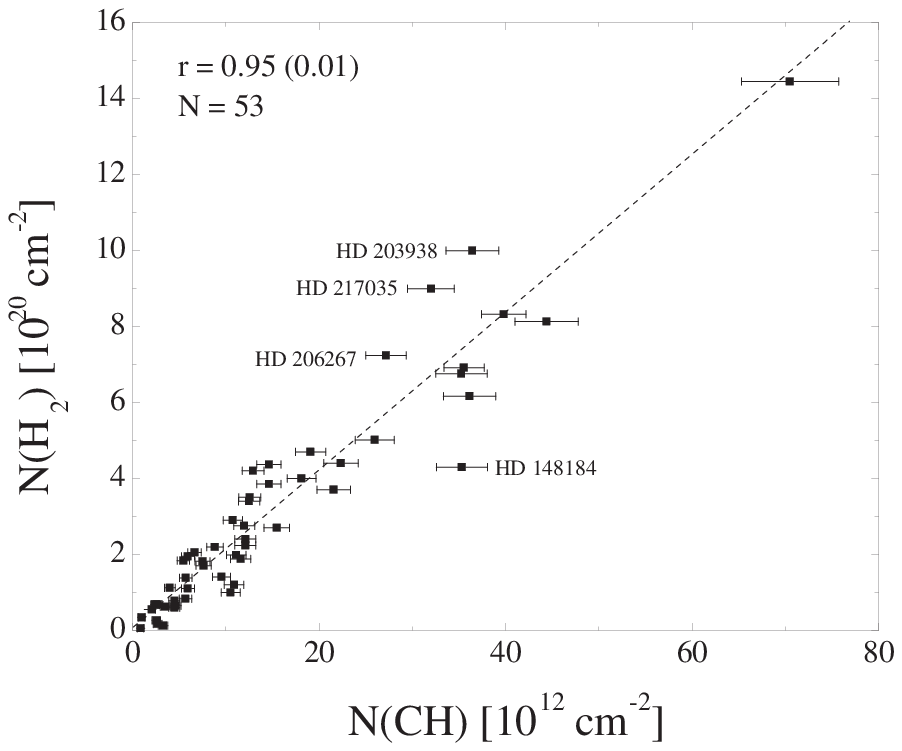}
\includegraphics[width=6.5 cm]{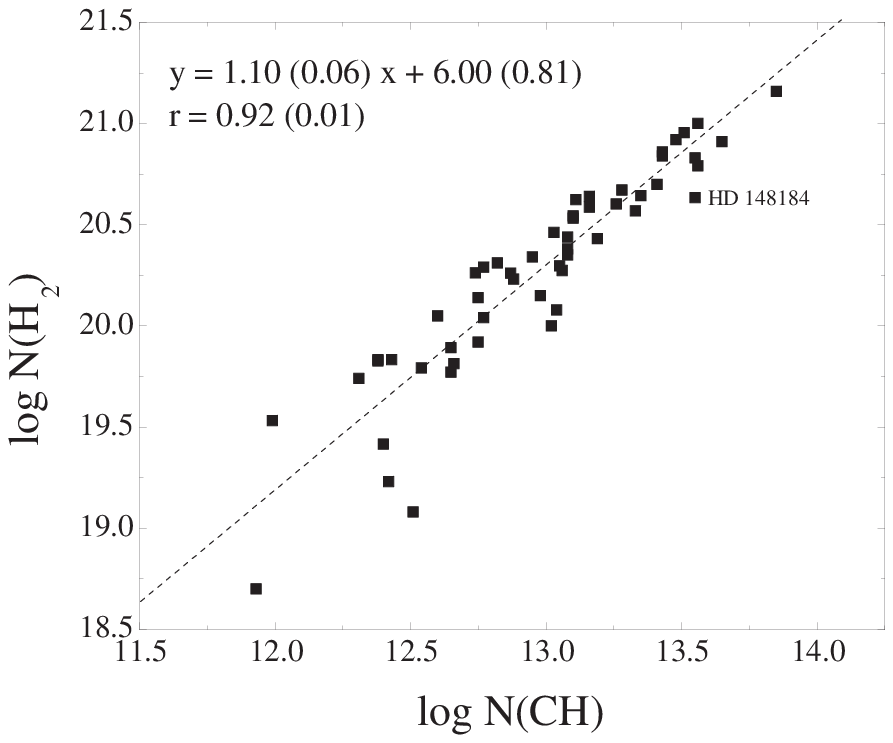}
\caption{  
Tight relation between column densities of CH and 
molecular hydrogen seen
also in normal and log scale (at the right panel). 
The correlation coefficient in each case is equal to 0.95 and
0.92, respectively.}

\label{fig}
\end{figure*}

\section{Acknowledgements}
\label{section3}
Authors acknowledge the financial support: JK and TW
acknowledge that of the Polish State during the period
2007 - 2010 (grant N203 012 32/1550).
We are grateful to Y. Byaletski., G. LoCurto, G.A. Galazutdiov 
and F.A. Musaev for their help in ackquiring the spectra used and many 
helpful suggestions.

\end{document}